\documentclass[twocolumn,pra,showpacs]{revtex4-2}
\usepackage{epsfig}
\usepackage{dcolumn}
\usepackage{natbib}
\usepackage[utf8]{inputenc}
\usepackage[english]{babel}
\usepackage{xcolor}
\usepackage{graphicx}
\usepackage{color}
\usepackage{bm}
\usepackage{amssymb}
\usepackage[intlimits]{amsmath}
\usepackage{booktabs}
\usepackage{amsmath}
\usepackage[para,online,flushleft]{threeparttable}

\definecolor{darkcandyapplered}{rgb}{0.64, 0.0, 0.0}
\usepackage[colorlinks=true,allcolors=darkcandyapplered]{hyperref}
\usepackage{physics}
\usepackage{multirow}
\begin{document}
\draft
\title{Variation of the quadrupole hyperfine structure and nuclear radius due to an interaction with scalar and axion dark matter}
\author{V. V. Flambaum} 
\email{v.flambaum@unsw.edu.au}
\author{A. J. Mansour}
\email{andrew.mansour@student.unsw.edu.au}
\address{School of Physics, University of New South Wales,
Sydney 2052, Australia}
\date{\today}

\begin{abstract}
Atomic spectroscopy is used to search for the space-time variation of fundamental constants which may be due to an interaction with scalar and pseudoscalar (axion) dark matter. In this letter, we study the effects which are produced by the variation of the nuclear radius and electric quadrupole moment. The sensitivity of the electric quadrupole hyperfine structure to both the variation of the quark mass and the effects of dark matter exceeds that of the magnetic hyperfine structure by 1-2 orders of magnitude. Therefore, the measurement of the variation of the ratio of the electric quadrupole and magnetic dipole hyperfine constants is proposed. The sensitivity of the optical clock transitions in the Yb$^+$ ion to the variation of the nuclear radius allows us to extract, from experimental data, limits on the variation of the hadron and quark masses, the QCD parameter $\theta$ and the interaction with axion dark matter. 
%
\end{abstract}

\maketitle

\textit{Introduction}. -- The present paper has two aims. The  first aim is to extract new limits on temporal variation of the fundamental constants and interactions of dark matter from existing atomic spectra measurements. The second aim is to propose new experiments where these effects of the variation and dark matter interactions are strongly enhanced. As intermediate steps we consider variation of the fundamental constants due to interaction with dark matter fields, dependence  of nuclear radius on variation of hadron parameters and effects of variation of nuclear radius on atomic spectra.  Note that all effects considered in this paper are related to the variation of the nuclear radius. 


One of the most important unsolved problems in physics is uncovering the nature of dark matter. Amongst other things, it is hypothesised that dark matter is made up of light bosonic particles, which are not accounted for in the standard model of elementary particles. The candidate particles in this class are the pseudoscalar axion (and axion like particles) and the dilaton-like scalar particle~\cite{Preskill,Abbott,Dine}. If the mass of the cold dark matter is very light ($m_{\text{DM}} \ll 1 \ \text{eV}$), it may be considered to be a classical field oscillating harmonically at every particular point in space. For axions, we may write this as

\begin{align}
a = a_{0} \cos (\omega t + \varphi), \ \omega \approx m_{a} \,,
\end{align}
where $\varphi$ is a (position-dependent) phase and $m_{a}$ is the mass of the axion. Assuming that axions saturate the entire dark matter density, the amplitude $a_{0}$ may be expressed in terms of the local dark matter density $\rho_{\text{DM}} \approx 0.4 \ \text{GeV/cm}^{3}$, see e.g. Ref.~\cite{DMdensity},

\begin{align}
a_{0} = \frac{\sqrt{2 \rho_{\text{DM}}}}{m_{a}} \,.
\end{align}
Similar expressions are used to describe the case of a scalar field dark matter. \\








\textit{Variation of fundamental constants due to interaction with  dark matter fields}. -- The effects of the interaction between scalar field dark matter and fermions 
may be presented as the apparent variation of fermion masses. This immediately follows from a comparison of the interaction of a fermion with the scalar field $ -g_f M_f  \phi^n \bar \psi \psi $ and the fermion mass term in the Lagrangian $- M_f \bar \psi \psi$. Adding these terms gives $M'_f=M_f(1+ g_f  \phi^n)$, $n=1,2$. Similarly, the interaction of scalar dark matter with the electromagnetic field may be accounted for as a variable fine structure constant $\alpha'=\alpha (1 + g_{\gamma} \phi^n)$, see, e.g., Refs.~\cite{Arvanitaki,Stadnik}. Note that variation of quark and electron masses and variation of $\alpha$ are determined by different interaction constants and may be treated as independent effects.

 The dependence of atomic transition frequencies on $\alpha$ and the quark masses  was calculated in  Refs.~\cite{PRLWebb,PRAWebb,CanJPh,Tedesco2006,Borschevsky,csquarks}. Atomic spectroscopy methods have already allowed one to place improved limits on the interaction strength of the low mass scalar field $\phi$ with photons, electrons and quarks by up to 15 orders in magnitude \cite{Stadnik,Stadnik2}. The experimental results have been obtained by the measurements of the oscillating frequency ratios of electron transitions in Dy/Cs \cite{DyCs}, Rb/Cs \cite{Hees2016}, Yb/Cs \cite{YbCs}, Sr/H/Si cavity \cite{HSi}, Cs/cavity \cite{Tretiak},  Yb$^+$/Yb$^+$/Sr \cite{Banerjee2023,Filzinger2023}.
 
 Note that if the interaction is quadratic in $\phi$, we may replace the scalar field by the pseudoscalar (axion) field as $\phi^2$ always has positive parity \cite{Stadnik}. The corresponding theory has been developed in Ref. \cite{Perez2022}, in which limits on the axion interaction from atomic spectroscopy experiments were obtained (see also \cite{Samsonov}). \\



\textit{Dependence of the nuclear radius on hadron and quark masses}. -- 
Ref.~\cite{Banerjee2023} proposed that the dependence of the electronic atomic transition frequencies on the nuclear radius (and subsequently on the hadronic parameters) may be used in the search for dark matter fields. Previously, in Ref. \cite{Dinh2009}, the dependence on the nuclear radius and hadronic parameters was studied in hyperfine transitions with the aim of searching for the variation of the fundamental constants.

Calculations of the dependence of nuclear energy levels and nuclear radii on fundamental constants were performed in Refs.~\cite{Wiringa2007,Wiringa2009}. Specifically, in Table VI of Ref.~\cite{Wiringa2009}, the sensitivity coefficients of nuclear radii to the variation of hadron masses for several light nuclei have been presented. These results may be extended to all nuclei. This is due to the fact that nuclear radii in all nuclei are quite accurately described by a universal formula $R_N=A^{1/3} r_0$, so in fact it is sufficient to calculate the dependence of $r_0$ in any nucleus.
The sensitivity coefficients are defined by the relation
\begin{align}
    \frac{\delta r_{0}}{r_{0}} = \sum_h K_{h} \frac{\delta m_{h}}{m_{h}} \,.
\end{align}
The sum over hadrons in Refs.~\cite{Wiringa2007,Wiringa2009}
includes  contributions of $\pi$, nucleon, $\Delta$ and vector mesons. The sensitivity to the pion mass is given by the coefficient $K_{\pi} = 1.8$ and the sensitivity to the nucleon mass is given by $K_{N} = -4.8$. 

Subsequently, the variation of hadron masses may be related to variation of the quark mass, see e.g.  Ref.~\cite{Roberts2008}:
\begin{equation}
  \frac{\delta m_h}{m_h} = K_{h,q} \frac{\delta m_q}{m_q}\,,
\end{equation}
where $m_{q} = (m_{u} + m_{d})/2$ corresponds to the average light quark mass. The sensitivity coefficient for the  pion mass is an order of magnitude bigger than that for other hadrons since the pion mass vanishes for zero quark mass ($m_{\pi} \propto m_q^{1/2}$) while other hadron masses remain finite. Indeed, according to Ref.~\cite{Roberts2008} for the pion  $K_{\pi,q}=0.498$, while for nucleons $K_{N,q}= 0.06$. The  sensitivity coefficients to the quark mass for light nuclei have been calculated in Ref.~\cite{Wiringa2009}. The average value is given by 
\begin{align} \label{dr}
    \frac{\delta r_{0}}{r_{0}} = 0.3 \frac{\delta m_{q}}{m_{q}} \,.
\end{align}
Note that here there are partial cancellations of different contributions, so the sensitivity is smaller than that following from pion mass alone. Refs.~\cite{Wiringa2007,Wiringa2009} have also presented calculations of the dependence of the nuclear energies and radii on variation of the fine structure constant $\alpha$.  \\

\textit{Limits on the drift of the nuclear radius and quark mass}. -- Now, using these results, along with the measurement of the drift of ratio of optical transition frequencies in  Yb$^+$ clock transitions from Ref.~\cite{Filzinger2023}, we can extract limits on the drift of the quark mass. This experiment measured the ratio of the $^{2}S_{1/2} (F = 0) \leftrightarrow \ ^{2} F_{7/2} (F=3)$ electric octupole (E3) and $^{2}S_{1/2} (F = 0) \leftrightarrow \ ^{2} D_{3/2} (F=2)$ electric quadrupole (E2) transition frequencies, and was used to measure the drift of the fine structure constant $\alpha$. However, due to the dependence of atomic transition frequencies on the nuclear radius their results are also sensitive to any variation of the quark mass, see Eq. (\ref{dr}). 

The dependence of atomic transition frequencies on the nuclear radius has been calculated for many atomic transitions with the aim to find isotope shifts. For E2 transitions in Yb$^+$ such calculations have been performed in Refs.~\cite{Counts,Allehabi}. For E3 transitions the dependence may be found using the measured ratio of isotope shifts for E3 and E2 transitions~\cite{Hur}. The result of such calculations is in excellent agreement with the result of the calculation of the sensitivity of the ratio of the E3 and E2 Yb$^+$ transition frequencies to the variation of the nuclear radius from Ref.~\cite{Banerjee2023},
\begin{align} 
\frac{\delta (\nu_{E3}/\nu_{E2})}{\nu_{E3}/\nu_{E2}}=2.4 \cdot 10^{-3} \frac{\delta \ev{r_n^2}}{\ev{r_n^2}}\,.
\end{align}

Using the measurements of the drift of atomic transition frequencies from Ref.~\cite{Filzinger2023},
\begin{align} 
\frac{\delta (\nu_{E3}/\nu_{E2})}{\nu_{E3}/\nu_{E2}}
=-1.2 (1.8) \cdot 10^{-18} \ \text{yr}^{-1} \,,
\end{align}
we obtain 
\begin{align} \label{nudrift}
 \frac{\delta \ev{r_n^2}}{\ev{r_n^2}}=-0.50(0.75) \cdot 10^{-15} \ \text{yr}^{-1} \,.
\end{align}
Using Eq. (\ref{dr}), we obtain the variation of the quark mass as
\begin{align} 
 \frac{\delta m_q}{m_q}=-0.83(1.25) \cdot 10^{-15} \ \text{yr}^{-1} \,.
\end{align}
This result is an improvement compared to the best limit obtained from measurements of the Cs/Rb hyperfine transition frequency ratio, $7.1(4.4) \cdot 10^{-15}\,\text{yr}^{-1}$, presented in Eq. (3) and in the fit of all available limits in Table III of Ref.~\cite{Guena}.

Note that when discussing the variation of dimensionful parameters, we should show the units we measure them in as units can also vary. For example, the SI units of frequency and time are defined by the Cs atom hyperfine structure constant which has a complicated dependence on the fundamental constants - see Eq. (\ref{Mhfs}).
In other words, we should consider the variation of dimensionless parameters which do not depend on any measurement units.  Nuclear properties depend on the quark mass and $\Lambda_{\text{QCD}}$. As we keep $\Lambda_{\text{QCD}}$ constant, we may say that we measure the variation of the dimensionless parameter $X_q=m_q/\Lambda_{\text{QCD}}$, i.e. we measure the quark mass in units of $\Lambda_{\text{QCD}}$ - see Refs. \cite{Wiringa2009,Tedesco2006}. A similar choice of units is assumed for the variation of hadron masses, considered below. 
\\ 

\textit{Limits on the drift of the hadron masses and the QCD parameter $\theta$}. -- For some applications, such as considering the limit on variation of the QCD parameter $\theta$, it is convenient to consider the problem at the hadron level, without going to the quark level.  
In Ref.~\cite{Wiringa2009}, the sensitivity of the nuclear radius to the masses of the pion, nucleon, vector meson and delta has been calculated. In the following estimate, we do not include contributions from the vector meson and delta as their contributions are smaller. These contributions also have opposing signs, meaning they partially cancel each other out making their contribution less reliable. The variation of the nuclear radius may be written in terms of the pion and nucleon mass as
\begin{align} \label{rpi}
 \frac{\delta r_0}{r_0}=1.8  \frac{\delta m_{\pi}}{m_{\pi}} -  4.8 \frac{\delta m_N}{m_N} = 1.2   \frac{\delta m_{\pi}}{m_{\pi}} \,,
\end{align}
where in the last equality we have used the following result from Ref.~\cite{Perez2022} 
\begin{align}\label{mppi} 
  \frac{\delta m_N}{m_N} = 0.13   \frac{\delta m_{\pi}}{m_{\pi}} \,.
\end{align}
From Eqs. (\ref{nudrift},\ref{rpi}) we can obtain a limit on the drift of the pion mass
\begin{align} \label{pidrift}
  \frac{\delta m_{\pi}}{m_{\pi}}  =-0.21(0.31) \cdot 10^{-15} \ \text{yr}^{-1} \,.
\end{align}
The pion mass depends on  the QCD parameter $\theta$. The shift  of the pion mass due to a small  $\theta$  relative to the pion mass for $\theta=0$ is given  by~\cite{Ubaldi2010}: 
\begin{align} \label{pitheta}
  \frac{\delta m_{\pi}}{m_{\pi}}  =-0.05 \theta^2 \,.
\end{align}
Using Eqs. (\ref{pidrift},\ref{pitheta})  we obtain constrains on the linear drift of $\theta^2$:
\begin{equation} 
  \frac{d\theta^2}{dt}  = 4(6) \cdot 10^{-15} \ \text{yr}^{-1}
\end{equation}

\textit{Limit on the interaction with the axion dark matter field}. -- 
Standard model spinor fields $\psi$, photon $F_{\mu\nu}$ and gluon $G^l_{\mu\nu}$ fields can have the following interaction vertices with a pseudoscalar field $a$:
\begin{equation}
\label{V1}
V= \frac{C_f}{f_a}  \partial_{\mu}  a  \bar \psi  \gamma_5 \gamma^{\mu}\psi + 
 C_{\gamma} \frac{a}{f_a} \tilde{F}^{\mu\nu} F_{\mu\nu} +  C_g \frac{a}{f_a} \tilde{G}^{l\,\mu\nu} G^l_{\mu\nu}\,.  
\end{equation}
Here $C_f$, $C_\gamma$ and $C_g$ are some dimensionless constants which are of order $O(1)$ for the QCD axion model, but are arbitrary for the general pseudoscalar (axion-like) particle. In particular, upon the substitution $C_g = g^2/(32\pi^2)$, or
\begin{equation}
\label{theta}
    \theta = \frac{32\pi^2 C_g a}{g^2 f_a}\,
\end{equation}
the last term in Eq.~(\ref{V1}) reduces 
to 
the standard QCD $\theta$-term
\begin{equation}
\label{thetaQCD}
    \frac{g^2\theta}{32\pi^2}\tilde G^{l\,\mu\nu}G^l_{\mu\nu}\,,
\end{equation}
where $\theta=a/f_a$, $f_a$ is the axion decay constant, $g$ is the strong interaction coupling constant, $G^l_{\mu\nu}$ is the gluon field strength and $\tilde G^{l\,\mu\nu}$ is its dual.
Thus, the classical axion dark matter field $a=a_0 \cos (m_a t + \varphi)$ may be interpreted as a dynamical QCD parameter $\theta=a/f_a$
~\cite{Preskill,Abbott,Dine}. 

According to Ref. \cite{Banerjee2023}, the measurement of the oscillation of the ratio of frequencies $\nu_{E3}$ and $\nu_{E2}$ in Yb$^+$ clock transitions may be used to study the axion dark matter field $a=a_0 \cos (m_a t+ \varphi)$. This is due to the dependence of the nuclear radius on $\theta$.
However, our result for the dependence of the nuclear radius on the pion mass Eq. (\ref{rpi}) is 6 times bigger in magnitude and has a different sign: our coefficient is $\beta=1.2$ while the coefficient in Ref.~\cite{Banerjee2023} is $\beta=-0.2$. Therefore, according to our calculations, the limits on the axion interaction should be 6 times stronger than those presented on the exclusion plot on Fig. 2 of Ref.~\cite{Banerjee2023}.

Note that the sign of $\beta$ may be determined without calculation. An increase in the pion mass leads to a decrease of the interaction range, i.e. a decrease of the effect of pion exchange potentials and a decrease of nuclear binding which leads to an increase in the nuclear radius. Thus, $\beta$ must be positive. The difference in sign is due to the assumption in Ref. \cite{Banerjee2023} that the nuclear radius is proportional to the nucleon radius. However, the internucleon distance $r_0$  is actually determined by the position of the minimum of the internucleon potential. Similarly, the sign of the coefficient -4.8 describing the dependence of $r_0$ on the nucleon mass  $m_p$  is explained by the decrease of kinetic energy $p^2/2 m_p$, increase of binding energy and decrease of $r_0$ if the nucleon mass increases. \\


\textit{Variation of the nuclear electric quadrupole moment and quadrupole hyperfine structure due to scalar and axion dark matter.} -- The variation of the nuclear radius also leads to the variation of the nuclear electric quadrupole moment $Q$ and quadrupole hyperfine structure constant $B$ which are proportional to $r_0^2$. Using Eq. (\ref{dr}) we obtain the following for the effect of the quark mass, with $ K_q=0.3$
\begin{equation} \label{Bq}
\frac{\delta B}{B}  =0.6 \frac{\delta m_q}{m_q} \,.
\end{equation}
We see that the sensitivity of the quadrupole constant $B$ to quark mass variation is 1-2 orders of magnitude higher than the sensitivity of the magnetic hyperfine constant $A$ calculated in~\cite{Tedesco2006}. Indeed, the dependence of the magnetic moment on quark mass in Cs is given by $K_q=0.009$, whilst in Rb it is given by $K_q= -0.016$~\cite{Tedesco2006}. One may measure the variation of the ratio of $B/A$ in the same atom  and achieve a significant improvement in the sensitivity to the variation of the quark mass, $\theta$ and the interaction with the scalar/axion dark matter field. The variation of the magnetic hyperfine constant ratio in Cs and Rb and the corresponding effect of the scalar dark matter field have been measured in Refs. \cite{Guena,Hees2016}. 

Let us start by stating the dependence of the electric quadrupole hyperfine constant $B$ on the fundamental constants:
\begin{align} 
B \propto  \frac{e Q}{a^3} \sim \frac{e^2 r_0^2 }{a_{B}^3} \,,
\end{align}
where $a_{B}$ is the Bohr radius. The dependence of the magnetic hyperfine constant is given by the following expression:
\begin{align}\label{Mhfs} 
A \propto  \frac{\mu_B \mu_N}{a_{B}^3} \sim \frac{ e^2 \hbar^2 g_N}{m_e m_p c^2 a_{B}^3} \,,
\end{align}
where  $\mu_B$ and  $\mu_N$ are the Bohr and nuclear magnetons and $g_N$ is the nuclear magnetic $g$-factor. Therefore, the ratio of the electric quadrupole and magnetic dipole constants may be written as  
\begin{align} \label{BA}
\frac{B}{A} \propto  \frac{ r_0^2 m_e m_p c^2 R_B(Z\alpha)}{g_N \hbar^2 R_A(Z \alpha)}. 
\end{align}
Here, we have added the relativistic factors $R_B(Z\alpha)$  and  $R_A(Z \alpha)$ for the electric quadrupole and magnetic dipole constants respectively, which are presented e.g in the paper~\cite{Tedesco2006} and the book~\cite{Sobelman}. These factors may be of interest if one searches for the variation of the fine structure constant $\alpha$. Oscillations of $\alpha$ and the electron and quark masses may be due to linear or quadratic interaction of the scalar field $\phi=\phi_0 \cos (m_{\phi}t + \varphi) $ with photons, electrons and quarks~\cite{Arvanitaki,Stadnik}. Therefore, measurements of the ratio $B/A$ may be used to search for scalar dark matter (and  axion dark matter in the case of interaction with $\phi^2$).

The dependence of the magnetic hyperfine constant $A$ on hadron parameters is different for different nuclei and rather weak~\cite{Tedesco2006,Perez2022}. Thus, in performing a zeroth order approximation we may neglect this dependence and present general estimates which are valid for all nuclei. The dominating effect comes from the variation of $r_0^2$ in Eq. (\ref{BA}). 
Using Eqs. (\ref{dr},\ref{rpi},\ref{pitheta},\ref{Bq}), we obtain 
\begin{align} \label{BAtheta}
\frac{\delta (B/A)}{B/A} \approx 0.6 \frac{\delta m_q}{m_q}\approx 2.4   \frac{\delta m_{\pi}}{m_{\pi}} \approx  -0.12 \theta^2 \,. 
\end{align}
Substituting $\theta=a/f_a$, we see that measurements of $B/A$ may be used to search for the axion dark matter field $a = a_{0} \cos (m_a t + \varphi)$.

Atoms and ions with nuclear spin $I>1/2$ in a state with electron angular momentum $J>1/2$ have both  electric quadrupole and magnetic dipole hyperifine interactions. In principle, any such systems are suitable for the measurements of $B/A$ variation, with the sensitivity defined by Eqs. (\ref{BA},\ref{BAtheta}).

An interesting possibility may be the measurement of the variation of the electric quadrupole hyperfine structure in diamagnetic polar molecules. In this case, the nuclear electric quadrupole moment interacts with the electric field of the polar molecule.  There is no electron angular momentum involved and no magnetic hyperfine structure.  This may reduce systematic effects. One can measure, for example, the variation of the ratio of the frequency of the transition between the components of electric quadrupole hyperfine structure in such molecules to the transition frequency in Cs or Rb clocks which is defined by the magnetic hyperfine constant.   A similar electric quadrupole interaction exists in solids where a large number of atoms reduces the statistical error. \\

\textit{Summary.} -- 
Atomic spectroscopy allows one to search for the space-time variation of fundamental constants and low mass scalar and pseudoscalar (axion) dark matter fields, which may be a source of such variation. 
The effects of varying fundamental constants include a variation in the nuclear radius.
One method of placing constraints on the variation in the nuclear radius is via a measurement of the variation in optical clock transition frequencies. 
Our calculation of the sensitivity of the Yb$^{+}$ transition frequencies to the variation of the nuclear radius agrees with that from Ref.~\cite{Banerjee2023}. Using  the measurements of the drift of atomic transition frequencies from Ref.~\cite{Filzinger2023}, we place constraints on the variation of the nuclear radius. We use this result and nuclear calculations to significantly improve limits on the variation of the quark masses. We then obtain limits on the variation of the QCD parameter $\theta$ and on the interaction with axion and scalar dark matter. 

Any variation in the nuclear radius leads to a variation in the nuclear electric quadrupole moment, and thus a variation in the quadrupole hyperfine structure constant $B$. The sensitivity of the quadrupole constant $B$ to the variation of quark masses and to  the interaction with dark matter is 1-2 orders of magnitude higher than the sensitivity of the magnetic hyperfine constant $A$, considered in previous publications. This implies that one may measure the variation of the ratio $B/A$ in the same atom, and achieve a significant improvement in the sensitivity to the variation of the quark mass, $\theta$ and the interaction with the scalar/axion dark matter fields.  As such, we estimate the dependence of $B/A$ on these quantities. One can also measure variation of the ratio of the frequency of the transition between the components of electric quadrupole hyperfine structure in a diamagnetic  molecule to the transition frequency in Cs or Rb clocks which is defined by the magnetic hyperfine constant. \\

\textit{Acknowledgements.} -- This work was supported by the Australian Research Council Grants No.\ DP230101058 and DP200100150.

\bibliographystyle{apsrev4-2}
\bibliography{References.bib}

\begin{thebibliography}{32}%
\makeatletter
\providecommand \@ifxundefined [1]{%
 \@ifx{#1\undefined}
}%
\providecommand \@ifnum [1]{%
 \ifnum #1\expandafter \@firstoftwo
 \else \expandafter \@secondoftwo
 \fi
}%
\providecommand \@ifx [1]{%
 \ifx #1\expandafter \@firstoftwo
 \else \expandafter \@secondoftwo
 \fi
}%
\providecommand \natexlab [1]{#1}%
\providecommand \enquote  [1]{``#1''}%
\providecommand \bibnamefont  [1]{#1}%
\providecommand \bibfnamefont [1]{#1}%
\providecommand \citenamefont [1]{#1}%
\providecommand \href@noop [0]{\@secondoftwo}%
\providecommand \href [0]{\begingroup \@sanitize@url \@href}%
\providecommand \@href[1]{\@@startlink{#1}\@@href}%
\providecommand \@@href[1]{\endgroup#1\@@endlink}%
\providecommand \@sanitize@url [0]{\catcode `\\12\catcode `\$12\catcode
  `\&12\catcode `\#12\catcode `\^12\catcode `\_12\catcode `\%12\relax}%
\providecommand \@@startlink[1]{}%
\providecommand \@@endlink[0]{}%
\providecommand \url  [0]{\begingroup\@sanitize@url \@url }%
\providecommand \@url [1]{\endgroup\@href {#1}{\urlprefix }}%
\providecommand \urlprefix  [0]{URL }%
\providecommand \Eprint [0]{\href }%
\providecommand \doibase [0]{https://doi.org/}%
\providecommand \selectlanguage [0]{\@gobble}%
\providecommand \bibinfo  [0]{\@secondoftwo}%
\providecommand \bibfield  [0]{\@secondoftwo}%
\providecommand \translation [1]{[#1]}%
\providecommand \BibitemOpen [0]{}%
\providecommand \bibitemStop [0]{}%
\providecommand \bibitemNoStop [0]{.\EOS\space}%
\providecommand \EOS [0]{\spacefactor3000\relax}%
\providecommand \BibitemShut  [1]{\csname bibitem#1\endcsname}%
\let\auto@bib@innerbib\@empty
\bibitem [{\citenamefont {Preskill}\ \emph {et~al.}(1983)\citenamefont
  {Preskill}, \citenamefont {Wise},\ and\ \citenamefont {Wilczek}}]{Preskill}%
  \BibitemOpen
  \bibfield  {author} {\bibinfo {author} {\bibfnamefont {J.}~\bibnamefont
  {Preskill}}, \bibinfo {author} {\bibfnamefont {M.~B.}\ \bibnamefont {Wise}},\
  and\ \bibinfo {author} {\bibfnamefont {F.}~\bibnamefont {Wilczek}},\ }\href
  {https://doi.org/https://doi.org/10.1016/0370-2693(83)90637-8} {\bibfield
  {journal} {\bibinfo  {journal} {Phys. Lett. B}\ }\textbf {\bibinfo {volume}
  {120}},\ \bibinfo {pages} {127} (\bibinfo {year} {1983})}\BibitemShut
  {NoStop}%
\bibitem [{\citenamefont {Abbott}\ and\ \citenamefont
  {Sikivie}(1983)}]{Abbott}%
  \BibitemOpen
  \bibfield  {author} {\bibinfo {author} {\bibfnamefont {L.}~\bibnamefont
  {Abbott}}\ and\ \bibinfo {author} {\bibfnamefont {P.}~\bibnamefont
  {Sikivie}},\ }\href
  {https://doi.org/https://doi.org/10.1016/0370-2693(83)90638-X} {\bibfield
  {journal} {\bibinfo  {journal} {Phys. Lett. B}\ }\textbf {\bibinfo {volume}
  {120}},\ \bibinfo {pages} {133} (\bibinfo {year} {1983})}\BibitemShut
  {NoStop}%
\bibitem [{\citenamefont {Dine}\ and\ \citenamefont {Fischler}(1983)}]{Dine}%
  \BibitemOpen
  \bibfield  {author} {\bibinfo {author} {\bibfnamefont {M.}~\bibnamefont
  {Dine}}\ and\ \bibinfo {author} {\bibfnamefont {W.}~\bibnamefont
  {Fischler}},\ }\href
  {https://doi.org/https://doi.org/10.1016/0370-2693(83)90639-1} {\bibfield
  {journal} {\bibinfo  {journal} {Phys. Lett. B}\ }\textbf {\bibinfo {volume}
  {120}},\ \bibinfo {pages} {137} (\bibinfo {year} {1983})}\BibitemShut
  {NoStop}%
\bibitem [{\citenamefont {Read}(2014)}]{DMdensity}%
  \BibitemOpen
  \bibfield  {author} {\bibinfo {author} {\bibfnamefont {J.~I.}\ \bibnamefont
  {Read}},\ }\href {https://doi.org/10.1088/0954-3899/41/6/063101} {\bibfield
  {journal} {\bibinfo  {journal} {J. Phys. G}\ }\textbf {\bibinfo {volume}
  {41}},\ \bibinfo {pages} {063101} (\bibinfo {year} {2014})}\BibitemShut
  {NoStop}%
\bibitem [{\citenamefont {Arvanitaki}\ \emph {et~al.}(2015)\citenamefont
  {Arvanitaki}, \citenamefont {Huang},\ and\ \citenamefont
  {Van~Tilburg}}]{Arvanitaki}%
  \BibitemOpen
  \bibfield  {author} {\bibinfo {author} {\bibfnamefont {A.}~\bibnamefont
  {Arvanitaki}}, \bibinfo {author} {\bibfnamefont {J.}~\bibnamefont {Huang}},\
  and\ \bibinfo {author} {\bibfnamefont {K.}~\bibnamefont {Van~Tilburg}},\
  }\href {https://doi.org/10.1103/PhysRevD.91.015015} {\bibfield  {journal}
  {\bibinfo  {journal} {Phys. Rev. D}\ }\textbf {\bibinfo {volume} {91}},\
  \bibinfo {pages} {015015} (\bibinfo {year} {2015})}\BibitemShut {NoStop}%
\bibitem [{\citenamefont {Stadnik}\ and\ \citenamefont
  {Flambaum}(2015)}]{Stadnik}%
  \BibitemOpen
  \bibfield  {author} {\bibinfo {author} {\bibfnamefont {Y.~V.}\ \bibnamefont
  {Stadnik}}\ and\ \bibinfo {author} {\bibfnamefont {V.~V.}\ \bibnamefont
  {Flambaum}},\ }\href {https://doi.org/10.1103/PhysRevLett.115.201301}
  {\bibfield  {journal} {\bibinfo  {journal} {Phys. Rev. Lett.}\ }\textbf
  {\bibinfo {volume} {115}},\ \bibinfo {pages} {201301} (\bibinfo {year}
  {2015})}\BibitemShut {NoStop}%
\bibitem [{\citenamefont {Dzuba}\ \emph
  {et~al.}(1999{\natexlab{a}})\citenamefont {Dzuba}, \citenamefont {Flambaum},\
  and\ \citenamefont {Webb}}]{PRLWebb}%
  \BibitemOpen
  \bibfield  {author} {\bibinfo {author} {\bibfnamefont {V.~A.}\ \bibnamefont
  {Dzuba}}, \bibinfo {author} {\bibfnamefont {V.~V.}\ \bibnamefont
  {Flambaum}},\ and\ \bibinfo {author} {\bibfnamefont {J.~K.}\ \bibnamefont
  {Webb}},\ }\href {https://doi.org/10.1103/PhysRevLett.82.888} {\bibfield
  {journal} {\bibinfo  {journal} {Phys. Rev. Lett.}\ }\textbf {\bibinfo
  {volume} {82}},\ \bibinfo {pages} {888} (\bibinfo {year}
  {1999}{\natexlab{a}})}\BibitemShut {NoStop}%
\bibitem [{\citenamefont {Dzuba}\ \emph
  {et~al.}(1999{\natexlab{b}})\citenamefont {Dzuba}, \citenamefont {Flambaum},\
  and\ \citenamefont {Webb}}]{PRAWebb}%
  \BibitemOpen
  \bibfield  {author} {\bibinfo {author} {\bibfnamefont {V.~A.}\ \bibnamefont
  {Dzuba}}, \bibinfo {author} {\bibfnamefont {V.~V.}\ \bibnamefont
  {Flambaum}},\ and\ \bibinfo {author} {\bibfnamefont {J.~K.}\ \bibnamefont
  {Webb}},\ }\href {https://doi.org/10.1103/PhysRevA.59.230} {\bibfield
  {journal} {\bibinfo  {journal} {Phys. Rev. A}\ }\textbf {\bibinfo {volume}
  {59}},\ \bibinfo {pages} {230} (\bibinfo {year}
  {1999}{\natexlab{b}})}\BibitemShut {NoStop}%
\bibitem [{\citenamefont {Flambaum}\ and\ \citenamefont
  {Dzuba}(2009)}]{CanJPh}%
  \BibitemOpen
  \bibfield  {author} {\bibinfo {author} {\bibfnamefont {V.~V.}\ \bibnamefont
  {Flambaum}}\ and\ \bibinfo {author} {\bibfnamefont {V.~A.}\ \bibnamefont
  {Dzuba}},\ }\href {https://doi.org/10.1139/p08-072} {\bibfield  {journal}
  {\bibinfo  {journal} {Can. J. Phys.}\ }\textbf {\bibinfo {volume} {87}},\
  \bibinfo {pages} {25} (\bibinfo {year} {2009})}\BibitemShut {NoStop}%
\bibitem [{\citenamefont {Flambaum}\ and\ \citenamefont
  {Tedesco}(2006)}]{Tedesco2006}%
  \BibitemOpen
  \bibfield  {author} {\bibinfo {author} {\bibfnamefont {V.~V.}\ \bibnamefont
  {Flambaum}}\ and\ \bibinfo {author} {\bibfnamefont {A.~F.}\ \bibnamefont
  {Tedesco}},\ }\href {https://doi.org/10.1103/PhysRevC.73.055501} {\bibfield
  {journal} {\bibinfo  {journal} {Phys. Rev. C}\ }\textbf {\bibinfo {volume}
  {73}},\ \bibinfo {pages} {055501} (\bibinfo {year} {2006})}\BibitemShut
  {NoStop}%
\bibitem [{\citenamefont {Pa\v{s}teka}\ \emph {et~al.}(2019)\citenamefont
  {Pa\v{s}teka}, \citenamefont {Hao}, \citenamefont {Borschevsky},
  \citenamefont {Flambaum},\ and\ \citenamefont {Schwerdtfeger}}]{Borschevsky}%
  \BibitemOpen
  \bibfield  {author} {\bibinfo {author} {\bibfnamefont {L.~F.}\ \bibnamefont
  {Pa\v{s}teka}}, \bibinfo {author} {\bibfnamefont {Y.}~\bibnamefont {Hao}},
  \bibinfo {author} {\bibfnamefont {A.}~\bibnamefont {Borschevsky}}, \bibinfo
  {author} {\bibfnamefont {V.~V.}\ \bibnamefont {Flambaum}},\ and\ \bibinfo
  {author} {\bibfnamefont {P.}~\bibnamefont {Schwerdtfeger}},\ }\href
  {https://doi.org/10.1103/PhysRevLett.122.160801} {\bibfield  {journal}
  {\bibinfo  {journal} {Phys. Rev. Lett.}\ }\textbf {\bibinfo {volume} {122}},\
  \bibinfo {pages} {160801} (\bibinfo {year} {2019})}\BibitemShut {NoStop}%
\bibitem [{\citenamefont {Flambaum}\ and\ \citenamefont
  {Munro-Laylim}(2023)}]{csquarks}%
  \BibitemOpen
  \bibfield  {author} {\bibinfo {author} {\bibfnamefont {V.~V.}\ \bibnamefont
  {Flambaum}}\ and\ \bibinfo {author} {\bibfnamefont {P.}~\bibnamefont
  {Munro-Laylim}},\ }\href {https://doi.org/10.1103/PhysRevD.107.015004}
  {\bibfield  {journal} {\bibinfo  {journal} {Phys. Rev. D}\ }\textbf {\bibinfo
  {volume} {107}},\ \bibinfo {pages} {015004} (\bibinfo {year}
  {2023})}\BibitemShut {NoStop}%
\bibitem [{\citenamefont {Stadnik}\ and\ \citenamefont
  {Flambaum}(2016)}]{Stadnik2}%
  \BibitemOpen
  \bibfield  {author} {\bibinfo {author} {\bibfnamefont {Y.~V.}\ \bibnamefont
  {Stadnik}}\ and\ \bibinfo {author} {\bibfnamefont {V.~V.}\ \bibnamefont
  {Flambaum}},\ }\href {https://doi.org/10.1103/PhysRevA.94.022111} {\bibfield
  {journal} {\bibinfo  {journal} {Phys. Rev. A}\ }\textbf {\bibinfo {volume}
  {94}},\ \bibinfo {pages} {022111} (\bibinfo {year} {2016})}\BibitemShut
  {NoStop}%
\bibitem [{\citenamefont {Van~Tilburg}\ \emph {et~al.}(2015)\citenamefont
  {Van~Tilburg}, \citenamefont {Leefer}, \citenamefont {Bougas},\ and\
  \citenamefont {Budker}}]{DyCs}%
  \BibitemOpen
  \bibfield  {author} {\bibinfo {author} {\bibfnamefont {K.}~\bibnamefont
  {Van~Tilburg}}, \bibinfo {author} {\bibfnamefont {N.}~\bibnamefont {Leefer}},
  \bibinfo {author} {\bibfnamefont {L.}~\bibnamefont {Bougas}},\ and\ \bibinfo
  {author} {\bibfnamefont {D.}~\bibnamefont {Budker}},\ }\href
  {https://doi.org/10.1103/PhysRevLett.115.011802} {\bibfield  {journal}
  {\bibinfo  {journal} {Phys. Rev. Lett.}\ }\textbf {\bibinfo {volume} {115}},\
  \bibinfo {pages} {011802} (\bibinfo {year} {2015})}\BibitemShut {NoStop}%
\bibitem [{\citenamefont {Hees}\ \emph {et~al.}(2016)\citenamefont {Hees},
  \citenamefont {Gu\'ena}, \citenamefont {Abgrall}, \citenamefont {Bize},\ and\
  \citenamefont {Wolf}}]{Hees2016}%
  \BibitemOpen
  \bibfield  {author} {\bibinfo {author} {\bibfnamefont {A.}~\bibnamefont
  {Hees}}, \bibinfo {author} {\bibfnamefont {J.}~\bibnamefont {Gu\'ena}},
  \bibinfo {author} {\bibfnamefont {M.}~\bibnamefont {Abgrall}}, \bibinfo
  {author} {\bibfnamefont {S.}~\bibnamefont {Bize}},\ and\ \bibinfo {author}
  {\bibfnamefont {P.}~\bibnamefont {Wolf}},\ }\href
  {https://doi.org/10.1103/PhysRevLett.117.061301} {\bibfield  {journal}
  {\bibinfo  {journal} {Phys. Rev. Lett.}\ }\textbf {\bibinfo {volume} {117}},\
  \bibinfo {pages} {061301} (\bibinfo {year} {2016})}\BibitemShut {NoStop}%
\bibitem [{\citenamefont {Kobayashi}\ \emph {et~al.}(2022)\citenamefont
  {Kobayashi}, \citenamefont {Takamizawa}, \citenamefont {Akamatsu},
  \citenamefont {Kawasaki}, \citenamefont {Nishiyama}, \citenamefont {Hosaka},
  \citenamefont {Hisai}, \citenamefont {Wada}, \citenamefont {Inaba},
  \citenamefont {Tanabe},\ and\ \citenamefont {Yasuda}}]{YbCs}%
  \BibitemOpen
  \bibfield  {author} {\bibinfo {author} {\bibfnamefont {T.}~\bibnamefont
  {Kobayashi}}, \bibinfo {author} {\bibfnamefont {A.}~\bibnamefont
  {Takamizawa}}, \bibinfo {author} {\bibfnamefont {D.}~\bibnamefont
  {Akamatsu}}, \bibinfo {author} {\bibfnamefont {A.}~\bibnamefont {Kawasaki}},
  \bibinfo {author} {\bibfnamefont {A.}~\bibnamefont {Nishiyama}}, \bibinfo
  {author} {\bibfnamefont {K.}~\bibnamefont {Hosaka}}, \bibinfo {author}
  {\bibfnamefont {Y.}~\bibnamefont {Hisai}}, \bibinfo {author} {\bibfnamefont
  {M.}~\bibnamefont {Wada}}, \bibinfo {author} {\bibfnamefont {H.}~\bibnamefont
  {Inaba}}, \bibinfo {author} {\bibfnamefont {T.}~\bibnamefont {Tanabe}},\ and\
  \bibinfo {author} {\bibfnamefont {M.}~\bibnamefont {Yasuda}},\ }\href
  {https://doi.org/10.1103/PhysRevLett.129.241301} {\bibfield  {journal}
  {\bibinfo  {journal} {Phys. Rev. Lett.}\ }\textbf {\bibinfo {volume} {129}},\
  \bibinfo {pages} {241301} (\bibinfo {year} {2022})}\BibitemShut {NoStop}%
\bibitem [{\citenamefont {Kennedy}\ \emph {et~al.}(2020)\citenamefont
  {Kennedy}, \citenamefont {Oelker}, \citenamefont {Robinson}, \citenamefont
  {Bothwell}, \citenamefont {Kedar}, \citenamefont {Milner}, \citenamefont
  {Marti}, \citenamefont {Derevianko},\ and\ \citenamefont {Ye}}]{HSi}%
  \BibitemOpen
  \bibfield  {author} {\bibinfo {author} {\bibfnamefont {C.~J.}\ \bibnamefont
  {Kennedy}}, \bibinfo {author} {\bibfnamefont {E.}~\bibnamefont {Oelker}},
  \bibinfo {author} {\bibfnamefont {J.~M.}\ \bibnamefont {Robinson}}, \bibinfo
  {author} {\bibfnamefont {T.}~\bibnamefont {Bothwell}}, \bibinfo {author}
  {\bibfnamefont {D.}~\bibnamefont {Kedar}}, \bibinfo {author} {\bibfnamefont
  {W.~R.}\ \bibnamefont {Milner}}, \bibinfo {author} {\bibfnamefont {G.~E.}\
  \bibnamefont {Marti}}, \bibinfo {author} {\bibfnamefont {A.}~\bibnamefont
  {Derevianko}},\ and\ \bibinfo {author} {\bibfnamefont {J.}~\bibnamefont
  {Ye}},\ }\href {https://doi.org/10.1103/PhysRevLett.125.201302} {\bibfield
  {journal} {\bibinfo  {journal} {Phys. Rev. Lett.}\ }\textbf {\bibinfo
  {volume} {125}},\ \bibinfo {pages} {201302} (\bibinfo {year}
  {2020})}\BibitemShut {NoStop}%
\bibitem [{\citenamefont {Tretiak}\ \emph {et~al.}(2022)\citenamefont
  {Tretiak}, \citenamefont {Zhang}, \citenamefont {Figueroa}, \citenamefont
  {Antypas}, \citenamefont {Brogna}, \citenamefont {Banerjee}, \citenamefont
  {Perez},\ and\ \citenamefont {Budker}}]{Tretiak}%
  \BibitemOpen
  \bibfield  {author} {\bibinfo {author} {\bibfnamefont {O.}~\bibnamefont
  {Tretiak}}, \bibinfo {author} {\bibfnamefont {X.}~\bibnamefont {Zhang}},
  \bibinfo {author} {\bibfnamefont {N.~L.}\ \bibnamefont {Figueroa}}, \bibinfo
  {author} {\bibfnamefont {D.}~\bibnamefont {Antypas}}, \bibinfo {author}
  {\bibfnamefont {A.}~\bibnamefont {Brogna}}, \bibinfo {author} {\bibfnamefont
  {A.}~\bibnamefont {Banerjee}}, \bibinfo {author} {\bibfnamefont
  {G.}~\bibnamefont {Perez}},\ and\ \bibinfo {author} {\bibfnamefont
  {D.}~\bibnamefont {Budker}},\ }\href
  {https://doi.org/10.1103/PhysRevLett.129.031301} {\bibfield  {journal}
  {\bibinfo  {journal} {Phys. Rev. Lett.}\ }\textbf {\bibinfo {volume} {129}},\
  \bibinfo {pages} {031301} (\bibinfo {year} {2022})}\BibitemShut {NoStop}%
\bibitem [{\citenamefont {Banerjee}\ \emph {et~al.}(2023)\citenamefont
  {Banerjee}, \citenamefont {Budker}, \citenamefont {Filzinger}, \citenamefont
  {Huntemann}, \citenamefont {Paz}, \citenamefont {Perez}, \citenamefont
  {Porsev},\ and\ \citenamefont {Safronova}}]{Banerjee2023}%
  \BibitemOpen
  \bibfield  {author} {\bibinfo {author} {\bibfnamefont {A.}~\bibnamefont
  {Banerjee}}, \bibinfo {author} {\bibfnamefont {D.}~\bibnamefont {Budker}},
  \bibinfo {author} {\bibfnamefont {M.}~\bibnamefont {Filzinger}}, \bibinfo
  {author} {\bibfnamefont {N.}~\bibnamefont {Huntemann}}, \bibinfo {author}
  {\bibfnamefont {G.}~\bibnamefont {Paz}}, \bibinfo {author} {\bibfnamefont
  {G.}~\bibnamefont {Perez}}, \bibinfo {author} {\bibfnamefont
  {S.}~\bibnamefont {Porsev}},\ and\ \bibinfo {author} {\bibfnamefont
  {M.}~\bibnamefont {Safronova}},\ }\href@noop {} {\bibinfo {title}
  {Oscillating nuclear charge radii as sensors for ultralight dark matter}}
  (\bibinfo {year} {2023}),\ \Eprint {https://arxiv.org/abs/2301.10784}
  {arXiv:2301.10784 [hep-ph]} \BibitemShut {NoStop}%
\bibitem [{\citenamefont {Filzinger}\ \emph {et~al.}(2023)\citenamefont
  {Filzinger}, \citenamefont {Dörscher}, \citenamefont {Lange}, \citenamefont
  {Klose}, \citenamefont {Steinel}, \citenamefont {Benkler}, \citenamefont
  {Peik}, \citenamefont {Lisdat},\ and\ \citenamefont
  {Huntemann}}]{Filzinger2023}%
  \BibitemOpen
  \bibfield  {author} {\bibinfo {author} {\bibfnamefont {M.}~\bibnamefont
  {Filzinger}}, \bibinfo {author} {\bibfnamefont {S.}~\bibnamefont
  {Dörscher}}, \bibinfo {author} {\bibfnamefont {R.}~\bibnamefont {Lange}},
  \bibinfo {author} {\bibfnamefont {J.}~\bibnamefont {Klose}}, \bibinfo
  {author} {\bibfnamefont {M.}~\bibnamefont {Steinel}}, \bibinfo {author}
  {\bibfnamefont {E.}~\bibnamefont {Benkler}}, \bibinfo {author} {\bibfnamefont
  {E.}~\bibnamefont {Peik}}, \bibinfo {author} {\bibfnamefont {C.}~\bibnamefont
  {Lisdat}},\ and\ \bibinfo {author} {\bibfnamefont {N.}~\bibnamefont
  {Huntemann}},\ }\href@noop {} {\bibinfo {title} {Improved limits on the
  coupling of ultralight bosonic dark matter to photons from optical atomic
  clock comparisons}} (\bibinfo {year} {2023}),\ \Eprint
  {https://arxiv.org/abs/2301.03433} {arXiv:2301.03433 [physics.atom-ph]}
  \BibitemShut {NoStop}%
\bibitem [{\citenamefont {Kim}\ and\ \citenamefont {Perez}(2022)}]{Perez2022}%
  \BibitemOpen
  \bibfield  {author} {\bibinfo {author} {\bibfnamefont {H.}~\bibnamefont
  {Kim}}\ and\ \bibinfo {author} {\bibfnamefont {G.}~\bibnamefont {Perez}}\
  }\href {https://doi.org/arXiv: 2205.12988} {arXiv: 2205.12988} (\bibinfo
  {year} {2022}),\ \Eprint {https://arxiv.org/abs/2205.12988} {arXiv:2205.12988
  [hep-ph]} \BibitemShut {NoStop}%
\bibitem [{\citenamefont {Flambaum}\ and\ \citenamefont
  {Samsonov}(2023)}]{Samsonov}%
  \BibitemOpen
  \bibfield  {author} {\bibinfo {author} {\bibfnamefont {V.~V.}\ \bibnamefont
  {Flambaum}}\ and\ \bibinfo {author} {\bibfnamefont {I.~B.}\ \bibnamefont
  {Samsonov}},\ }\href@noop {} {\bibinfo {title} {Fluctuations of atomic energy
  levels due to axion and scalar fields}} (\bibinfo {year} {2023}),\ \Eprint
  {https://arxiv.org/abs/2302.11167} {arXiv:2302.11167 [hep-ph]} \BibitemShut
  {NoStop}%
\bibitem [{\citenamefont {Dinh}\ \emph {et~al.}(2009)\citenamefont {Dinh},
  \citenamefont {Dunning}, \citenamefont {Dzuba},\ and\ \citenamefont
  {Flambaum}}]{Dinh2009}%
  \BibitemOpen
  \bibfield  {author} {\bibinfo {author} {\bibfnamefont {T.~H.}\ \bibnamefont
  {Dinh}}, \bibinfo {author} {\bibfnamefont {A.}~\bibnamefont {Dunning}},
  \bibinfo {author} {\bibfnamefont {V.~A.}\ \bibnamefont {Dzuba}},\ and\
  \bibinfo {author} {\bibfnamefont {V.~V.}\ \bibnamefont {Flambaum}},\ }\href
  {https://doi.org/10.1103/physreva.79.054102} {\bibfield  {journal} {\bibinfo
  {journal} {Phys. Rev. A}\ }\textbf {\bibinfo {volume} {79}},\ \bibinfo
  {pages} {054102} (\bibinfo {year} {2009})}\BibitemShut {NoStop}%
\bibitem [{\citenamefont {Flambaum}\ and\ \citenamefont
  {Wiringa}(2007)}]{Wiringa2007}%
  \BibitemOpen
  \bibfield  {author} {\bibinfo {author} {\bibfnamefont {V.~V.}\ \bibnamefont
  {Flambaum}}\ and\ \bibinfo {author} {\bibfnamefont {R.~B.}\ \bibnamefont
  {Wiringa}},\ }\href {https://doi.org/10.1103/PhysRevC.76.054002} {\bibfield
  {journal} {\bibinfo  {journal} {Phys. Rev. C}\ }\textbf {\bibinfo {volume}
  {76}},\ \bibinfo {pages} {054002} (\bibinfo {year} {2007})}\BibitemShut
  {NoStop}%
\bibitem [{\citenamefont {Flambaum}\ and\ \citenamefont
  {Wiringa}(2009)}]{Wiringa2009}%
  \BibitemOpen
  \bibfield  {author} {\bibinfo {author} {\bibfnamefont {V.~V.}\ \bibnamefont
  {Flambaum}}\ and\ \bibinfo {author} {\bibfnamefont {R.~B.}\ \bibnamefont
  {Wiringa}},\ }\href {https://doi.org/10.1103/PhysRevC.79.034302} {\bibfield
  {journal} {\bibinfo  {journal} {Phys. Rev. C}\ }\textbf {\bibinfo {volume}
  {79}},\ \bibinfo {pages} {034302} (\bibinfo {year} {2009})}\BibitemShut
  {NoStop}%
\bibitem [{\citenamefont {{Clo{\"e}t}}\ \emph {et~al.}(2008)\citenamefont
  {{Clo{\"e}t}}, \citenamefont {{Eichmann}}, \citenamefont {{Flambaum}},
  \citenamefont {{Roberts}}, \citenamefont {{Bhagwat}},\ and\ \citenamefont
  {{H{\"o}ll}}}]{Roberts2008}%
  \BibitemOpen
  \bibfield  {author} {\bibinfo {author} {\bibfnamefont {I.~C.}\ \bibnamefont
  {{Clo{\"e}t}}}, \bibinfo {author} {\bibfnamefont {G.}~\bibnamefont
  {{Eichmann}}}, \bibinfo {author} {\bibfnamefont {V.~V.}\ \bibnamefont
  {{Flambaum}}}, \bibinfo {author} {\bibfnamefont {C.~D.}\ \bibnamefont
  {{Roberts}}}, \bibinfo {author} {\bibfnamefont {M.~S.}\ \bibnamefont
  {{Bhagwat}}},\ and\ \bibinfo {author} {\bibfnamefont {A.}~\bibnamefont
  {{H{\"o}ll}}},\ }\href {https://doi.org/10.1007/s00601-008-0240-8} {\bibfield
   {journal} {\bibinfo  {journal} {Few-Body Systems}\ }\textbf {\bibinfo
  {volume} {42}},\ \bibinfo {pages} {91} (\bibinfo {year} {2008})},\ \Eprint
  {https://arxiv.org/abs/0804.3118} {arXiv:0804.3118 [nucl-th]} \BibitemShut
  {NoStop}%
\bibitem [{\citenamefont {Counts}\ \emph {et~al.}(2020)\citenamefont {Counts},
  \citenamefont {Hur}, \citenamefont {Aude~Craik}, \citenamefont {Jeon},
  \citenamefont {Leung}, \citenamefont {Berengut}, \citenamefont {Geddes},
  \citenamefont {Kawasaki}, \citenamefont {Jhe},\ and\ \citenamefont
  {Vuleti\ifmmode~\acute{c}\else \'{c}\fi{}}}]{Counts}%
  \BibitemOpen
  \bibfield  {author} {\bibinfo {author} {\bibfnamefont {I.}~\bibnamefont
  {Counts}}, \bibinfo {author} {\bibfnamefont {J.}~\bibnamefont {Hur}},
  \bibinfo {author} {\bibfnamefont {D.~P.~L.}\ \bibnamefont {Aude~Craik}},
  \bibinfo {author} {\bibfnamefont {H.}~\bibnamefont {Jeon}}, \bibinfo {author}
  {\bibfnamefont {C.}~\bibnamefont {Leung}}, \bibinfo {author} {\bibfnamefont
  {J.~C.}\ \bibnamefont {Berengut}}, \bibinfo {author} {\bibfnamefont
  {A.}~\bibnamefont {Geddes}}, \bibinfo {author} {\bibfnamefont
  {A.}~\bibnamefont {Kawasaki}}, \bibinfo {author} {\bibfnamefont
  {W.}~\bibnamefont {Jhe}},\ and\ \bibinfo {author} {\bibfnamefont
  {V.}~\bibnamefont {Vuleti\ifmmode~\acute{c}\else \'{c}\fi{}}},\ }\href
  {https://doi.org/10.1103/PhysRevLett.125.123002} {\bibfield  {journal}
  {\bibinfo  {journal} {Phys. Rev. Lett.}\ }\textbf {\bibinfo {volume} {125}},\
  \bibinfo {pages} {123002} (\bibinfo {year} {2020})}\BibitemShut {NoStop}%
\bibitem [{\citenamefont {Allehabi}\ \emph {et~al.}(2021)\citenamefont
  {Allehabi}, \citenamefont {Dzuba}, \citenamefont {Flambaum},\ and\
  \citenamefont {Afanasjev}}]{Allehabi}%
  \BibitemOpen
  \bibfield  {author} {\bibinfo {author} {\bibfnamefont {S.~O.}\ \bibnamefont
  {Allehabi}}, \bibinfo {author} {\bibfnamefont {V.~A.}\ \bibnamefont {Dzuba}},
  \bibinfo {author} {\bibfnamefont {V.~V.}\ \bibnamefont {Flambaum}},\ and\
  \bibinfo {author} {\bibfnamefont {A.~V.}\ \bibnamefont {Afanasjev}},\ }\href
  {https://doi.org/10.1103/PhysRevA.103.L030801} {\bibfield  {journal}
  {\bibinfo  {journal} {Phys. Rev. A}\ }\textbf {\bibinfo {volume} {103}},\
  \bibinfo {pages} {L030801} (\bibinfo {year} {2021})}\BibitemShut {NoStop}%
\bibitem [{\citenamefont {Hur}\ \emph {et~al.}(2022)\citenamefont {Hur},
  \citenamefont {Aude~Craik}, \citenamefont {Counts}, \citenamefont {Knyazev},
  \citenamefont {Caldwell}, \citenamefont {Leung}, \citenamefont {Pandey},
  \citenamefont {Berengut}, \citenamefont {Geddes}, \citenamefont {Nazarewicz},
  \citenamefont {Reinhard}, \citenamefont {Kawasaki}, \citenamefont {Jeon},
  \citenamefont {Jhe},\ and\ \citenamefont {Vuleti\ifmmode~\acute{c}\else
  \'{c}\fi{}}}]{Hur}%
  \BibitemOpen
  \bibfield  {author} {\bibinfo {author} {\bibfnamefont {J.}~\bibnamefont
  {Hur}}, \bibinfo {author} {\bibfnamefont {D.~P.~L.}\ \bibnamefont
  {Aude~Craik}}, \bibinfo {author} {\bibfnamefont {I.}~\bibnamefont {Counts}},
  \bibinfo {author} {\bibfnamefont {E.}~\bibnamefont {Knyazev}}, \bibinfo
  {author} {\bibfnamefont {L.}~\bibnamefont {Caldwell}}, \bibinfo {author}
  {\bibfnamefont {C.}~\bibnamefont {Leung}}, \bibinfo {author} {\bibfnamefont
  {S.}~\bibnamefont {Pandey}}, \bibinfo {author} {\bibfnamefont {J.~C.}\
  \bibnamefont {Berengut}}, \bibinfo {author} {\bibfnamefont {A.}~\bibnamefont
  {Geddes}}, \bibinfo {author} {\bibfnamefont {W.}~\bibnamefont {Nazarewicz}},
  \bibinfo {author} {\bibfnamefont {P.-G.}\ \bibnamefont {Reinhard}}, \bibinfo
  {author} {\bibfnamefont {A.}~\bibnamefont {Kawasaki}}, \bibinfo {author}
  {\bibfnamefont {H.}~\bibnamefont {Jeon}}, \bibinfo {author} {\bibfnamefont
  {W.}~\bibnamefont {Jhe}},\ and\ \bibinfo {author} {\bibfnamefont
  {V.}~\bibnamefont {Vuleti\ifmmode~\acute{c}\else \'{c}\fi{}}},\ }\href
  {https://doi.org/10.1103/PhysRevLett.128.163201} {\bibfield  {journal}
  {\bibinfo  {journal} {Phys. Rev. Lett.}\ }\textbf {\bibinfo {volume} {128}},\
  \bibinfo {pages} {163201} (\bibinfo {year} {2022})}\BibitemShut {NoStop}%
\bibitem [{\citenamefont {Gu\'ena}\ \emph {et~al.}(2012)\citenamefont
  {Gu\'ena}, \citenamefont {Abgrall}, \citenamefont {Rovera}, \citenamefont
  {Rosenbusch}, \citenamefont {Tobar}, \citenamefont {Laurent}, \citenamefont
  {Clairon},\ and\ \citenamefont {Bize}}]{Guena}%
  \BibitemOpen
  \bibfield  {author} {\bibinfo {author} {\bibfnamefont {J.}~\bibnamefont
  {Gu\'ena}}, \bibinfo {author} {\bibfnamefont {M.}~\bibnamefont {Abgrall}},
  \bibinfo {author} {\bibfnamefont {D.}~\bibnamefont {Rovera}}, \bibinfo
  {author} {\bibfnamefont {P.}~\bibnamefont {Rosenbusch}}, \bibinfo {author}
  {\bibfnamefont {M.~E.}\ \bibnamefont {Tobar}}, \bibinfo {author}
  {\bibfnamefont {P.}~\bibnamefont {Laurent}}, \bibinfo {author} {\bibfnamefont
  {A.}~\bibnamefont {Clairon}},\ and\ \bibinfo {author} {\bibfnamefont
  {S.}~\bibnamefont {Bize}},\ }\href
  {https://doi.org/10.1103/PhysRevLett.109.080801} {\bibfield  {journal}
  {\bibinfo  {journal} {Phys. Rev. Lett.}\ }\textbf {\bibinfo {volume} {109}},\
  \bibinfo {pages} {080801} (\bibinfo {year} {2012})}\BibitemShut {NoStop}%
\bibitem [{\citenamefont {Ubaldi}(2010)}]{Ubaldi2010}%
  \BibitemOpen
  \bibfield  {author} {\bibinfo {author} {\bibfnamefont {L.}~\bibnamefont
  {Ubaldi}},\ }\href {https://doi.org/10.1103/PhysRevD.81.025011} {\bibfield
  {journal} {\bibinfo  {journal} {Phys. Rev. D}\ }\textbf {\bibinfo {volume}
  {81}},\ \bibinfo {pages} {025011} (\bibinfo {year} {2010})}\BibitemShut
  {NoStop}%
\bibitem [{\citenamefont {Sobelman}(1979)}]{Sobelman}%
  \BibitemOpen
  \bibfield  {author} {\bibinfo {author} {\bibfnamefont {I.}~\bibnamefont
  {Sobelman}},\ }\href@noop {} {\emph {\bibinfo {title} {Atomic Spectra and
  Radiative Transitions}}}\ (\bibinfo  {publisher} {Springer-Verlag},\ \bibinfo
  {address} {Berlin Heidelberg New York},\ \bibinfo {year} {1979})\BibitemShut
  {NoStop}%
\end{thebibliography}%

\end{document}